\documentclass{emulateapj}\usepackage{apjfonts}

\usepackage{natbib}

\newcommand{\Msun}{M_{\odot}}
\newcommand{\Mearth}{Earth masses}

\slugcomment{Submitted to ApJ Letters, \today}
\shorttitle{Solar Twins, Solar and Jupiter abundances}
\shortauthors{Nordlund}

\begin{document}

\title{Solar Twins and Possible Solutions of the Solar and Jupiter Abundance Problems}

\author{\AA ke Nordlund}
\affil{Centre for Star and Planet Formation and Niels Bohr Institute,
University of Copenhagen, Juliane Maries Vej 30, DK-2100, Copenhagen, Denmark; aake@nbi.dk}

%-------------------------------------------------------------------------------
\begin{abstract}
Implications of the recently discovered systematic abundance difference between
the Sun and two collections of `solar twins' are discussed.  The differences
can be understood as an imprint on the abundances of the solar convection zone
caused by the lock-up of heavy elements in the planets.  Such a scenario also
leads naturally to possible solutions of two other abundance peculiarities; 1)
the discrepancy between photospheric abundances derived from accurate 3-D
models of the solar photosphere and the abundance of heavy elements in the
solar interior deduced from helioseismology, and 2) the abundance pattern of
Jupiter, which can either---with great difficulty---be interpreted as a general
and similar overabundance of both common elements such as carbon, nitrogen and
sulphur and rare inert gases such as argon, krypton and xenon, or---much more
simply---as an under-abundance of hydrogen.
\end{abstract}

\keywords{Sun: abundances -- Jupiter: abundances -- solar system: formation -- convection}

%-------------------------------------------------------------------------------
\section{Introduction}
%-------------------------------------------------------------------------------
\cite{melendez_peculiar_2009} have recently found systematic differences
between the abundances of heavy elements in the Sun and a number of nearby
`solar twins', selected from the Hipparcos catalogue.  The differences are
tightly correlated with the condensation temperatures of the elements
\citep{lodders_solar_2003}, and are very well established, with formal
probabilities of occurring in a random sample of the order of $10^{-9}$.
\cite{ramirez_2009} have found similar differences for a selection of northern
hemisphere solar twins. \cite{melendez_peculiar_2009} propose that the
abundance difference patterns could possibly be related to the presence of
planets.

One purpose of this Letter is to discuss this issue in some detail, and to
bring forward into the discussion some aspects that may not yet be generally
appreciated.  Among these is the fact that, when taking into account realistic
accretion histories from gravo-turbulent models of star formation
\cite{wuchterl_first_2001-1} find
that, contrary to conclusions from models based on a quasi-static picture
\citep[e.g.][]{dantona_new_1994}, the early Sun was not fully convective but
had a structure more or less homologous to the current Sun, with a convection
zone depth of only about one third of the radius
\citep{wuchterl_clouds_2003}.

Another aspect that, on closer inspection, is almost unavoidably related to this
issue is the so-called `solar abundance problem'
\citep[aka `the solar oxygen crisis';][]{ayres_solar_2008}.
Recent re-analyses of photospheric abundances of heavy
elements \citep{asplund_new_2005,caffau_photospheric_2008}; see also
\cite{asplund_2009} and \cite{nordlund_solar_2009},
lead to values of $Z_{\rm CZ}$, the mass fraction of elements heavier than
helium in the convection zone, between 0.012 and 0.015, while the best fit of
models of the solar interior to measurements of solar oscillations generally
suggest $Z\approx 0.017$ \citep[e.g.][]{antia_determining_2006}.
These latter model fits are mainly sensitive to the
abundances below the solar convection zone (CZ), while the photospheric
measurements of course pertain to the well-mixed solar CZ.  In principle these
abundances could differ, with the CZ value being smaller than the interior
value.
Since the present solar CZ contains about 2.5\% of the solar
mass, the `missing' amount of heavy elements in the CZ is only about 10--40 {\Mearth}
--- the exact amount depending on if one adopts the \cite{asplund_new_2005} or the
\cite{caffau_photospheric_2008} value of $Z_{\rm CZ}$.  This is comparable to
the mass of heavy elements locked up in the planets.  Thus, even
allowing for a somewhat more massive CZ in the early Sun, one would {\em
expect} to see a difference in CZ and interior heavy element abundance, of the
order of the one that is implied by the solar abundance problem.
This connection has been pointed out previously by \cite{haxton_cn_2008-1}.

A third issue that may very well also be related is the peculiar abundance
pattern of Jupiter \citep{owen_low-temperature_1999}.

In the following sections of this Letter I discuss, in turn, the likely initial
conditions for the collapse of the solar nebula (Section 2), the resulting
structure and early evolution of the Sun
(Section 3), and the conclusions one can draw with respect to the solar and
solar twin abundances on the background of the previous sections (Section 4).
Finally (Section 5) I discuss the abundance pattern of Jupiter in the light of
the previous conclusions.
The final Section summarizes the discussions and presents the overall
conclusions in compact form.

%-------------------------------------------------------------------------------
\section{Initial conditions for solar collapse}
%-------------------------------------------------------------------------------
A lot of experience has been accumulated over the last few years concerning the
conditions under which stars form in turbulent molecular clouds.  Numerical
simulations have been performed with both soft particle
hydrodynamics methods \citep[e.g.][]{klessen_quiescent_2005,bate_stellar_2009}
and grid methods \citep[e.g.][]{li_formation_2004,padoan_two_2007}.
A rather consistent picture has emerged, where it is understood that the
spectrum of initial masses of stars --- the {\em initial mass function} (IMF)
-- is the results of the interaction of turbulent fragmentation with
self-gravity, where supersonic and super-Alfv{\'e}nic turbulence chops up the
medium into a state with a very intermittent mass distribution
\citep{padoan_stellar_2002}.  A small fraction of the mass reaches, in
localized regions, sufficiently high densities to become gravitationally
unstable, and these regions then collapse to form stars.

In light of this highly dynamic picture of star formation one might well
question if the assumption of the near hydrostatic Bonnor-Ebert (B-E)
like initial state for solar collapse adopted by \cite{wuchterl_first_2001-1}
is realistic.  For reasons that are best explained with reference to what can
be gleaned from looking explicitly at how pre-stellar cores approach collapse
in simulations aimed at understanding star formation rates \citep{padoan_2009},
it is apparent that B-E like initial states are indeed reasonable approximations.

The power law part of the IMF distribution is a power law exactly because
essentially all local density maxima with that amount of mass reach densities
large enough for collapse.  In such cases
the collapse typically starts well before all mass that will eventually
converge has arrived; the collapse starts as soon as enough mass has arrived
for the local region to become gravitationally unstable, and additional mass
that arrives later on then just adds to an already established collapse.

The initial mass function starts to deviate significantly from the high
mass power law asymptote at masses of around 1--2 solar masses, because
some of these cores are unable to reach sufficient density for collapse.  Indeed,
the point where the IMF has a slope of -1, rather than a value close to the
Salpeter power law slope of {-1.35}, is near one solar mass, in empirical as well as
theoretical initial mass functions \citep{padoan_stellar_2002,chabrier_galactic_2003}.
A slope of -1 corresponds to a maximum in the contribution per unit logarithmic
mass interval from the IMF, so in this sense the Sun is a `typical star';
parcels of gas in star forming regions are more likely to end up in solar-like
stars than in more massive or less massive stars.

Irrespective of the differences in fractions of cores that collapse, and in
how early or late the collapse commences, the situations just before the
collapse are quite similar; mass is accumulating, with a density maximum
at some point is space, with nearby motions on the average converging relative
to the density maximum, thus systematically adding to the mass of
the growing core.  Eventually the core reaches a point in time where it becomes
gravitationally unstable, but just before that point in time it is a near-isothermal
structure, close to hydrostatic equilibrium, with self-gravity balanced by gas
(and magnetic) pressure \citep{ballesteros-paredes_molecular_2007}.
Such structures are necessarily similar to --- while
not identical to --- B-E spheres \citep{bonnor_stability_1958}.
Indeed, the similarity in structure between pre-stellar cores and BE-spheres
is well established observationally \citep{bacmann_isocam_2000}.

The choice by \cite{wuchterl_clouds_2003} to take BE-spheres as
initial conditions for models of stellar collapse is thus very well
motivated, and even though the detailed structure of pre-stellar cores --- in
particular in the outer parts --- may deviate somewhat from BE-spheres,
the inner and denser parts are likely to be well approximated by BE-spheres.

The assumption of one-dimensional accretion made by
\cite{wuchterl_first_2001-1} and \cite{wuchterl_clouds_2003} is a weakness. The
importance of this weakness is hard to estimate at the current time, since
there are no 2-D or 3-D simulation that also include an accurate representation
of the central star.  Most works that study the collapse of pre-stellar cores
in two- or three-dimensional focus on the question of fragmentation, and do not
attempt to study the interior structure of the resulting star.  The recent 2-D
model by \cite{tscharnuter_protostellar_2009-1} gives some hints, in that it
shows that a major part of the stellar mass rather quickly ends up in the
stellar embryo, after being processed in a relatively thick accretion structure
that surrounds the growing stellar embryo.  This indicates that the accretion
history and thermal evolution of the stellar embryo may turn out not be that
different when the full three-dimensional structure is included.

%-------------------------------------------------------------------------------
\section{Structure of the early Sun}
%-------------------------------------------------------------------------------
While many earlier models of the collapse of pre-stellar cores to stellar
densities have been presented \cite{wuchterl_first_2001-1} and
\cite{wuchterl_clouds_2003} appear to be the first to adopt a realistic
initial structure, and the first who have attempted to {\em predict} the
accretion history characteristic of the collapse of a solar mass core to
stellar densities, rather than just adopt some {\em ad hoc} accretion rate
prescription.

Subsequently \cite{froebrich_evolution_2006-1}
have gone a step further, predicting the typical distribution of luminosity
and effective temperature of Class 0 and Class I sources in star forming regions
by using accretion statistics derived from simulations of gravo-turbulent star
formation in molecular clouds.  The results are encouraging and provide much
better fits to observed distributions than models that assume parameterized,
constant accretion rates.  Similar efforts have
been undertaken by \cite{baraffe_episodic_2009-1}.

One of the most striking and significant differences between the
\cite{wuchterl_first_2001-1} models and other models of the early Sun is the
internal structure.  While earlier models predict an extended period of time
where the early Sun is fully convective, the \cite{wuchterl_first_2001-1} and
\cite{wuchterl_clouds_2003} models start out being essentially homologous to the
present Sun with, according to \cite{wuchterl_clouds_2003}, a convection zone
depth of only about one third of the radius.

This result has profound consequences for the interpretation of the abundance
pattern differences between the Sun and the solar twins, as well as for the
possible interpretation of the solar abundance problem in terms of an abundance
difference between the solar CZ and the solar interior \citep{haxton_cn_2008-1}.
It would thus be of great importance to have independent confirmation and / or
a more detailed analysis of the reasons for these fundamental differences in
internal structure.  For now, we take the results at face value, and go on to
discuss the implications.

%-------------------------------------------------------------------------------
\section{The Sun / solar twins abundance differences}
%-------------------------------------------------------------------------------
The abundance difference pattern between the Sun and most of the solar twins
reported by \cite{melendez_peculiar_2009} and \cite{ramirez_2009} is intriguing
and invites a search for an explanation.  \cite{melendez_peculiar_2009}
systematically examine several possibilities --- galactic evolution, supernova
pollution, and early dust separation --- but they conclude that that none of
them offer a compelling route to an explanation, and instead they settle for
element separation processes related to planet formation as the most likely
explanation. In short, the hypothesis is that the abundance difference pattern
between the Sun and most of the solar twins could find an explanation in the
more efficient `lock-up' of refractories in planets, as compared to volatiles.

The one major difficulty that \cite{melendez_peculiar_2009} find with this
explanation is that, given the traditional models where the early Sun was fully
convective \citep[e.g.][]{dantona_new_1994}, one would have to place the separation
processes at such a late time that it would require an unusually prolonged
dissipation time for the protoplanetary disk in the solar system.  However,
according to the discussion above, when taking the results of \cite{wuchterl_first_2001-1}
and \cite{wuchterl_clouds_2003} into account, the assumption of a delayed
evolution of the solar system is no longer needed.

%-------------------------------------------------------------------------------
\subsection{Heavy elements in the planets and in the solar CZ}
\label{sec:heavy}
%-------------------------------------------------------------------------------
With the problem of full mixing no longer on the table it is certainly worthwhile to
continue to examine the possibility that the elements locked up in planets have
left an observable imprint on solar abundances.  Indeed, with even just a
reasonably shallow convection zone in the early Sun, the situation is turned
completely up-side-down, in that --- as illustrated by the following discussion
--- it would then be surprising to {\em not} find such an effect.

The mass budgets are as follows:
The currently favored total heavy element contents of Jupiter and Saturn are
approximately 35 and 23 {\Mearth}, respectively \citep{saumon_shock_2004}.
The major reservoirs of hydrogen in the ice giants Uranus and Neptune are
probably their contents of water and ammonia, which means that by mass the
hydrogen content is rather negligible.  The helium content is harder to
estimate, but helium is never mentioned as a major constituent candidate of the
ice giants, so a reasonable estimate of the total mass of heavy elements in the
planets lands around 80 {\Mearth} (figures from 50 to 90 are mentioned in
the literature).

On the other hand the total mass of heavy elements in the current solar
convection zone is, assuming Z=0.012 \citep{asplund_new_2005} and $M_{\rm CZ} =
0.025 \Msun$, approximately 100 {\Mearth}. With similar amounts of heavy
elements locked up in planets, the only possibility to {\em not} see an effect
in solar convection zone abundances would be either 1) if the Sun was, after
all, fully convective (or nearly so) at the time of element separation by the
planets and accretion of the remnant gas,
or 2) if the hydrogen that corresponded to the heavy elements was more
or less completely ejected from the solar system. The first possibility
requires that a major mistake exists in the works of
\cite{wuchterl_first_2001-1} and \cite{wuchterl_clouds_2003}.  The possibility
to eject large amounts of primordial gas from the solar system through intense
external illumination has been examined by \cite{adams_early_2006}, who find
that it is unlikely with typical cluster conditions that such an effect
penetrates inside radii of about 30 AU.

We thus conclude that the heavy elements locked up in the planets are very likely
to have caused a significant abundance difference between the solar convection
zone and the solar interior, and since the existence of such a
difference could explain the apparent inconsistency between the observed
photospheric solar abundances and the solar interior ones deduced from
helioseismology, a hypothesis that such an abundance difference actually
exists and is the main cause of the solar abundance problem indeed is a
very attractive one, as advocated also by \cite{haxton_cn_2008-1}.

%-------------------------------------------------------------------------------
\subsection{The Sun / solar twins abundance difference pattern}
%-------------------------------------------------------------------------------
As per the discussion above it appears quite likely that the heavy elements
locked up in planets have left an imprint even on the total abundances of heavy
elements in the solar photosphere and convection zone.  But if that is the
case, we certainly should also expect to see differential effects, since it is
quite unlikely that all elements heavier than helium are present in solar
proportions in the planets.

The rocky planets are of course an illustration of this point, since they are
heavily overabundant in refractories relative to volatiles.  The fact that
meteorites (and hence asteroids)
show a differential abundance pattern very similar to the Sun /
solar twin pattern \citep{alexander_early_2001} is tantalizing as well, even
though asteroids represent a totally ignorable fraction of the solar system
mass.

The situation with respect to refractories and volatiles in the giant planets
is not known empirically, and in principle the differential effect of the
rocky planets could be neutralized by (or at least drowned in) the much larger
contributions from the giant planets.  It appears unlikely, however, that this
is the case, given for example the apparently diverse conditions under which
the gas and ice giant planets formed.  So, from a consideration of the rather
special circumstances that would be required to {\em not} have a differential
effect due the lock-up of heavy elements in planets, the presence of
differential effects between volatiles and refractories is not surprising.

Against this background the occurrence of relative abundance differences
between the Sun and solar twins is also not surprising by itself, since the
lock-up of heavy elements in planets would otherwise have to be very nearly the
same in the solar twins and in the Sun, to keep the differential effects below
the limits set by the abundance accuracies reached by
\cite{melendez_peculiar_2009} and \cite{ramirez_2009}.  The hypothesis
advocated by \cite{melendez_peculiar_2009}, that the abundance differences are
somehow related to the occurrence of planets, thus indeed seems very attractive.

It is difficult to judge from the relatively small number of cases investigated
until now if the differential abundance pattern is likely to be one between on
the one hand systems like the Sun \citep[there are 2-3 such systems in the 11 star
sample used by][]{melendez_peculiar_2009} and systems without planets, or if it
is one between systems that all have planets, but with differences in the
profiles of volatile and refractory element depletions.  \cite{melendez_peculiar_2009}
attempt to elucidate this question by comparing solar analogues with and without
known planets, but find the initially surprising result that it is the solar
analogues that do {\em not} have known planets that are more similar to the Sun,
while the ones that are known to have planets are more similar to the majority
of the solar twins.  Such a result can easily be explained, however, if the
(rather different) close-in gas giants typical of currently detected exo-planets
have a relatively more `flat' profile of volatile to refractory element
abundances.  As with the solar twins, a pattern similar to the Sun may be
taken to signal the presence of, as-of-yet undetected, more solar system like
planets.

Even though other conclusions cannot be excluded, it appears likely that the
observed abundance difference pattern, with its clear dependence on the
condensation temperature of particularly the refractory elements, is in fact
characteristic of the Sun's heavy element depletion due to the planets. If this
is indeed the case it is quite remarkable, in that the refractory
over-abundance pattern that is known to apply to just a very small fraction of
the solar system mass --- the meteorites and hence the asteroids --- would then
actually apply in essentially the same form to the whole collection of planets.

%-------------------------------------------------------------------------------
\section{The Jupiter abundance problem}
%-------------------------------------------------------------------------------
Jupiter has an abundance pattern that is from one point of view very peculiar.
Jupiter is the only planet in the solar system where hydrogen is the most common
chemical element (by number), and in that respect Jupiter is more akin to the
Sun than the other planets.  However, relative to the Sun Jupiter is still
overabundant in heavy elements.  In fact, a number of chemically very diverse
elements are overabundant with respect to the Sun by about the same factor of
three.  These elements include the volatile elements carbon, nitrogen and
sulphur, but also several of the inert gases; e.g., argon, krypton and xenon
\citep{owen_low-temperature_1999}.  While a number of other chemical elements (such as
oxygen among the volatiles and helium and neon among the inert gases) deviate
from the pattern, these deviations may well be due to specific interior processes
in Jupiter, as discussed by \cite{owen_low-temperature_1999}.

The fact that several rare and inert gases are enhanced by essentially the
same factor as some of the most common volatile elements is indeed a curious
circumstance, which is hard to explain in terms of enrichment of
Jupiter's atmosphere by accretion and capture of planetesimals, in that it
is very difficult to bind inert gases efficiently to planetesimals.  But,
lacking a better explanation \cite{owen_low-temperature_1999} nevertheless conclude that
"It seems to us that the only explanation ... is that these elements came to
the planet in very cold ($T < 30$ K) icy planetesimals", and a number of
subsequent and related papers have gone to great length to explain that such
bindings are possible via `clathration' at very low temperatures.

However, this entire line of explanations has a severe problem, irrespective
of how likely (or not) one finds a construction where inert gases are bound
to icy planetesimals at temperatures that probably occurred in the early solar
system only at very large distances from the Sun.  Even if the process is in
principle possible, it would have been totally unrelated to the
processes that are assumed to have enhanced the much more common elements.
Given that the processes are independent it would take a coincidence of some
proportions to have these processes produce virtually the same overabundance
(it may be unlikely even to have a population of more ordinary planetesimals
produce the same overabundance of carbon, nitrogen and sulphur).

On the other hand, in the context of a discussion where lock-up of elements in
planets appears to be an attractive possibility, another and much simpler
explanation offers itself: Jupiter's abundance pattern would be a natural
outcome of a situation where about 2/3 of the hydrogen that initially `belonged to'
Jupiter's heavy elements was lost into the Sun and never settled into Jupiter.
Rather than regarding Jupiter's abundance pattern as the result of an {\em
enhancement} of several heavy elements, one could also see it as the result of a {\em
depletion} of a single element; hydrogen.  Since all other planets are
depleted in hydrogen to a much more complete extent, such an explanation
seems {\em a priori} not at all unlikely.  A similar explanation has been
advanced by \cite{guillot_composition_2006}.

%-------------------------------------------------------------------------------
\section{Conclusions}
%-------------------------------------------------------------------------------
The newly discovered abundance pattern differences between the Sun and stars
(solar twins) that are in other respects virtually identical to the Sun
\citep{melendez_peculiar_2009,ramirez_2009} can possibly be understood as
an imprint on the abundances of the solar convection zone by the lock-up
of heavy elements in the planets.  Models of the formation of the Sun
that start out from pre-stellar core structures similar to those actually
observed result in early solar structures that are {\em not} fully convective
\citep{wuchterl_first_2001-1,wuchterl_clouds_2003}, and this removes the
need for peculiar assumptions such as delayed formation of the solar planets.
The credence of such a line of explanations is further strengthened by the
possibility to explain two other abundance related problems; 1) the apparent
discrepancy between increasingly accurate photospheric abundance determinations
\citep{asplund_2009}, and 2) the apparent overabundance of a number of chemically
very different element in Jupiters atmosphere, which can instead be interpreted
as an {\em under}-abundance of hydrogen, presumably stemming from the very
same process that, in the case of the other planets, was able to separate
hydrogen even more efficiently from the heavier elements.

\acknowledgements

I thank Bengt Gustafsson, Martin Bizzarro, David Arnett, and
Isabelle Baraffe for discussions and email exchanges.
This research was supported by a grant from the Danish Natural Science
Research Council (FNU). The Centre for Star and Planet Formation is
funded by the Danish National Research Foundation and the University
of Copenhagen's Programme of Excellence.

\bibliographystyle{apj}
\bibliography{adhoc,ms}

\end{document}